\documentclass[aps,pre,twocolumn,notitlepage,nofootinbib]{revtex4-1}
\usepackage{amsmath,amsfonts,amssymb,bm,cancel}
\usepackage{graphicx,float}
\usepackage{enumitem}
\usepackage{color}
\usepackage{soul}
\usepackage[dvipsnames]{xcolor}

\begin{document}

\title{\Large \textsf{Anomalous percolation transitions beyond the BKT transition in growing networks}}

\author{\large \textsf{S. M. Oh$^1$, S.-W. Son$^{2,3}$ and B. Kahng$^{1}$}}

\thanks{\textrm{$^1$CCSS, CTP and Department of Physics and Astronomy, Seoul National University, Seoul 08826, Korea, \\
$^2$Department of Applied Physics, Hanyang University, Ansan 15588, Korea, \\ $^3$Asia Pacific Center for Theoretical Physics, Pohang 37673, Korea \\e-mail: bkahng@snu.ac.kr }}

\maketitle

\textsf{{Since the discovery a half century ago that $1/r^2$-type long-range interactions in the one-dimensional Ising model change the phase transition type~\cite{thouless}, long-range interactions in diverse systems have received considerable attention. Recently, this interest extended to global suppression dynamics in the percolation transition, which changes a second-order transition to first order~\cite{warnke}. Here, we investigate how the Berezinskii-Kosterlitz-Thouless (BKT) transition is changed by the global suppression effect. In fact, this effect often arises in real-world complex systems, yet it is not appropriately accounted for in models. We find that the BKT transition breaks down, but the features of infinite-, second-, and first-order transitions all emerge as the link occupation probability is controlled. Moreover, we find that such growing networks exhibit maximum diversity, causing the mean cluster size to  diverge without formation of a giant cluster. We elucidate the underlying mechanisms and show that such anomalous transitions are universal.}} 

Berezinskii, Kosterlitz, and Thouless (BKT) discovered an infinite-order topological phase transition in early 1970s~\cite{bere_1971,bere_1972,kt_1972,kt_1973,kt_1974}. Since then, this type of transitions were observed in diverse phenomena ranging from the superfluid-normal phase transition~\cite{rmp_2017} and quantum phase transitions~\cite{rmp_quantum} in physical systems to percolation transitions (PTs) of growing networks~\cite{Callaway,mendes_BKT} in interdisciplinary areas. 

Growing networks are ubiquitous in the real world. Some examples are coauthorship networks~\cite{newman,coauthorship}, the World Wide Web (WWW)~\cite{www}, and protein interaction networks~\cite{sole,kim,vespignani}. In growing networks, the number of nodes increases with time. For instance, in the coauthorship network, as a new researcher joins in a research group, the network is growing. Callaway et al.~\cite{Callaway} introduced a simple model for such growing networks, called the growing random network (GRN) model: A node representing a person is present at the beginning. At each time, a new node is added in the system. A link is also added with probability $p$ between a pair of unconnected nodes chosen randomly among all existing nodes. As $p$ is increased, a PT occurs at the transition point $p_c$, beyond which a macroscopic-scale large cluster is generated. It was revealed that the PT of the GRN model follows the infinite-order BKT transition~\cite{Callaway,mendes_BKT,kim,sole}. The order parameter, the relative giant cluster size, $G(p)$ is zero for $p < p_c$, while it increases continuously for $p > p_c$ in the essentially singular form, $G(p)\sim  \exp(-a/\sqrt{p-p_c})$, where $a$ is a positive constant. The susceptibility, i.e., the mean cluster size $\langle s \rangle\equiv \sum_s s^2 n_s$ is finite both sides of a transition point. The behaviors of $G(p)$ and $\langle s \rangle$ are depicted schematically in Figs.~1{\bf a} and 1{\bf b}, respectively.

\begin{figure}[b]
\includegraphics[width=1.0\linewidth]{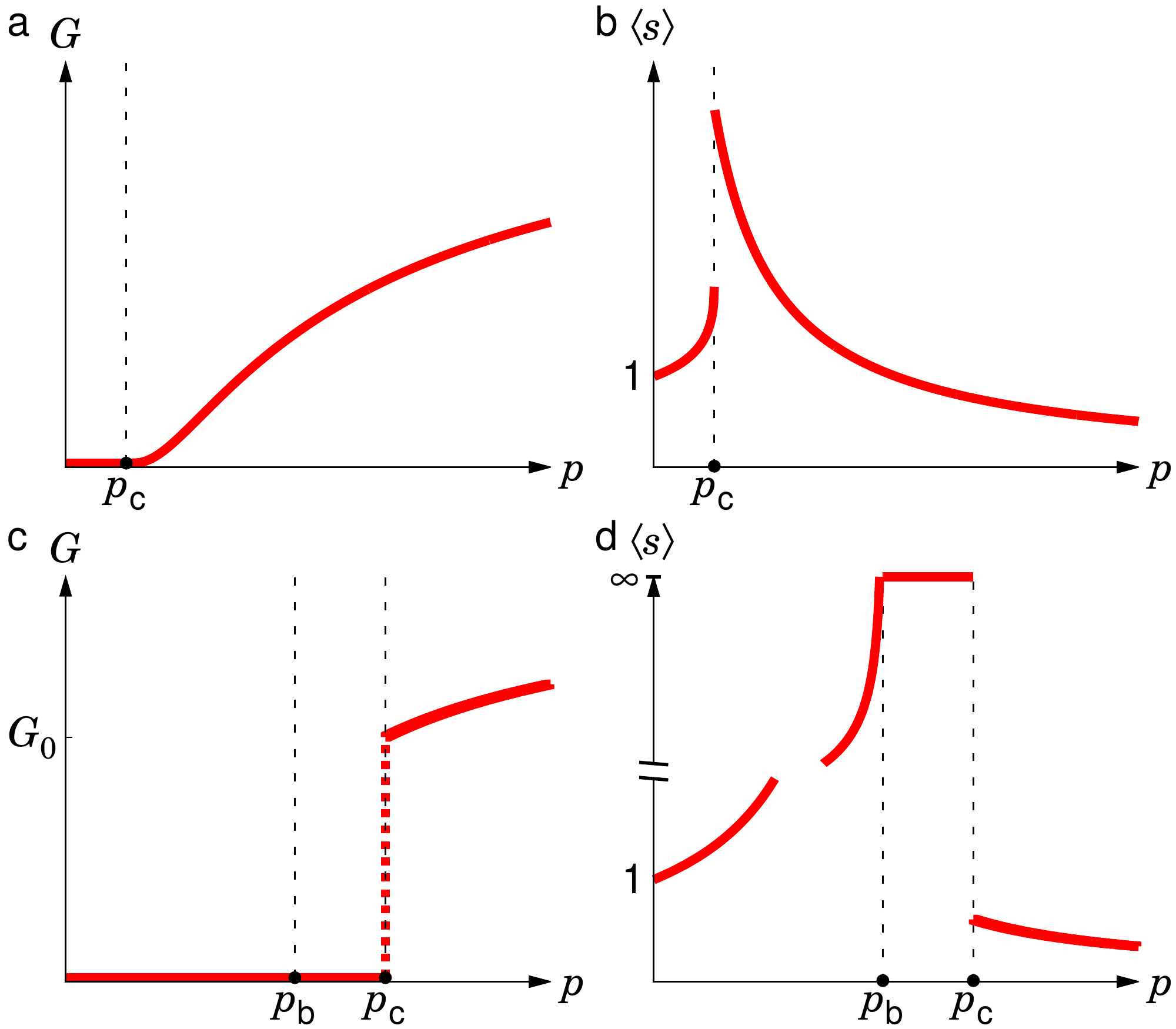}
\caption{{\bf Comparison of the order parameter and the susceptibility for the GRN and $r$-GRN models.} Plot of $G(p)$ and $\langle s \rangle$ versus $p$ for an infinite-order PT of the GRN model in {\bf a} and {\bf b}. Similar plot for a first-order transition of the $r$-GRN model in {\bf c} and {\bf d}.}
\label{fig:Callaway_vs_our_model}
\end{figure}

In statistical physics, it is well known that a phase transition type can be changed by long-range interactions~\cite{thouless}. For instance, the PT in one dimension is changed its type to an infinite-order transition by $1/r^2$ long-range connections~\cite{grassberger_BKT}. Similarly a PT can be changed by global suppression  dynamics~\cite{warnke}. For instance, when the formation of a spanning cluster is suppressed~\cite{avoiding}, the PT type can be changed from second order to first order. Such changes due to long-range connections or global suppression dynamics were considered in static networks; however, it has never been investigated yet in growing networks. Here, we aim to investigate how the infinite-order PT in growing networks is changed by the global suppression dynamics. 

Actually suppression dynamics may arise naturally in growing networks in the real world, for instance, the coauthorship network~\cite{coauthorship}. As a research group becomes larger, the group can become more inefficient functionally in some aspect; thus, new students are less likely to join such a large group and thus the growth of the large groups can be suppressed. As new students join small or medium groups, those groups grow in size. It was found empirically that the merging of large clusters occurs very rarely. Once it happens, it leads to an abrupt increase of a giant cluster~\cite{coauthorship}. Thus the evolution of the coauthorship network does not proceed by pure random connections, but there acts some suppression mechanism against the growth of large clusters. Moreover, the suppression effect can also arise in the WWW by the inaccessibility to a portal site. A global suppression dynamics may be seen in financial-ecological systems in the form of affirmative action policy. For instance, selected small enterprises are given more chances to promote their activities.

\begin{figure}[!ht]
\includegraphics[width=1.0\linewidth]{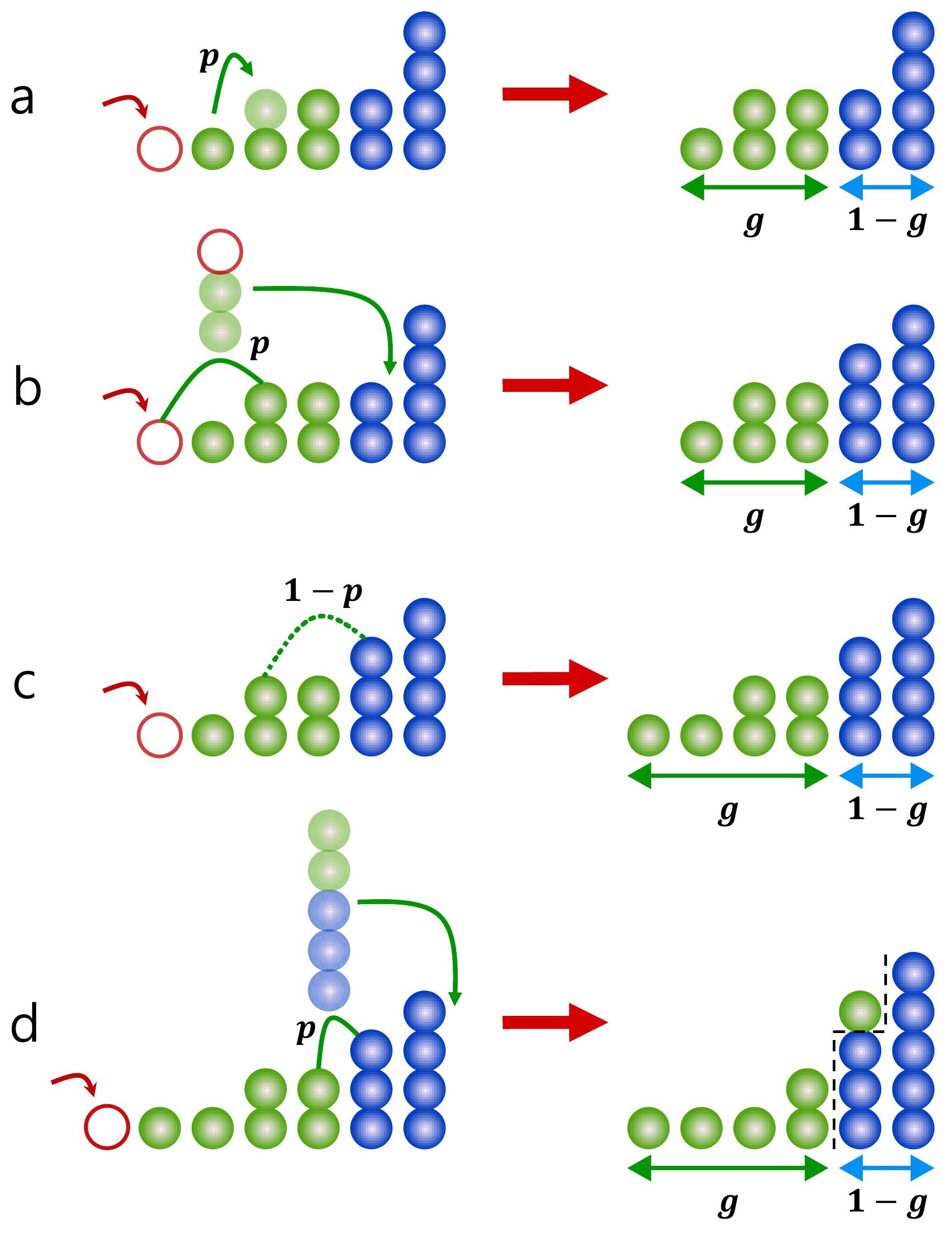}
\caption{{\bf Schematic illustration of the $r$-GRN model.} Nodes (represented by balls) in the set $R$ are dark green, whereas those in $R^c$ are blue. Each column represents a cluster. We start with the system at time $t=9$ that contains five clusters of sizes $(1, 1, 2, 2, 4)$ displayed from the second leftmost column to right in {\bf a}. Thus $N(9)=10$. Here a control parameter $g$ is taken as $g=0.4$, thus $\lceil gN \rceil=4$. The sets $R$ and $R^c$ contain four (dark green) and six (blue) nodes, respectively. {\bf a}. At time $t=10$, the leftmost red open node is newly added. $N(t)=11$ and $\lceil gN \rceil=5$. Next, two nodes, say, those belonging to the first and second dark green clusters of sizes ($1,1$) are selected, and they are merged with probability $p$, making a cluster of size two. The largest cluster size in $R$, denoted as $S_R$, remains two. {\bf b}. At time $t=11$, again a new node is added (open red ball). The new node is selected and merges into the second cluster in $R$, generating a cluster of size three. Because $N(11)=12$, $\lceil gN \rceil=5$. The merged cluster of size three moves to $R^c$, while the cluster of size two in $R^c$ moves to $R$. {\bf c}. At time $t=12$, a new node is added. $N(12)=13$ and $\lceil gN \rceil=6$. Two nodes are selected but they are not connected with probability $1-p$. {\bf d}. At time $t=13$, a new node is added. $N(13)=14$. $\lceil gN \rceil=6$. The fifth cluster from the left in $R$ and the first cluster in $R^c$ merge and generate a cluster of size five that belongs to $R^c$. The cluster of size four becomes lying on the border between the two sets. In this case, one node belongs to $R$, while the other three belong to $R^c$. $S_R$ is reset as four.}
\label{fig:Schematic_diagram}
\end{figure}

Here, to achieve our goal, we modify the GRN model by including a suppression dynamics as follows: At each time, a node is added to the system. To add a link, we select two nodes: a node from a portion of the smallest clusters and the other node from among all the nodes. They are connected with probability $p$. Because nodes belonging to the smallest clusters have twice the chance to be linked, while nodes in the remaining large clusters have single chance. The growth of large clusters is suppressed. The dynamic rule becomes global in the process of sorting out the portion of the smallest clusters among all cluster sizes at each time. This model is called the restricted growing random network ($r$-GRN) model hereafter. The detailed rule is presented schematically in Fig.~\ref{fig:Schematic_diagram} and will be described rigorously in the Method. On the other hand, a similar model was introduced for static networks, called the restricted Erd\H{o}s-R\'enyi ($r$-ER) model~\cite{ER_model,r_er,Hybrid_PT}. The globally suppression effect in the $r$-ER model changes the transition type from second order to hybrid.  

\begin{figure*}[t]
\includegraphics[width=1.0\linewidth]{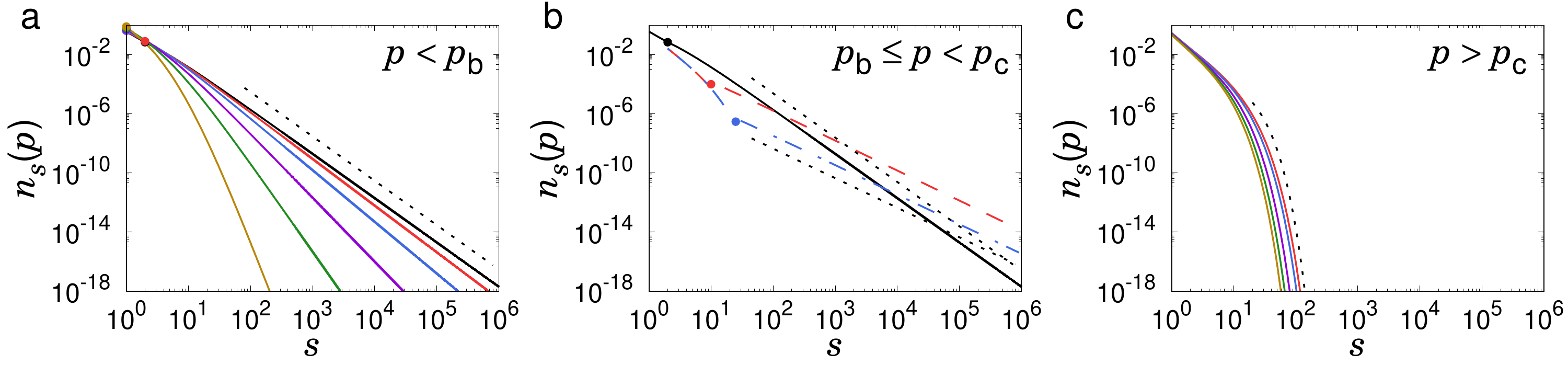}
\caption{{\bf The cluster size distribution $n_s(p)$ as a function of $s$ in different $p$ regions.} Three cases of $n_s(p)$ are distinguished for $g=0.4$: {\bf a}. For $p < p_b$, $n_s(p)$ asymptotically follows the power law $\sim s^{-\tau}$ with $\tau > 3$. The slope of the dotted guide line is $-3$. Solid lines are obtained for $p=0.472576 \approx p_{b}$, $0.45$, $0.4$, $0.3$, $0.2$, and $0.1$ from right to left. {\bf b}. For $p_b \leq p < p_c$, in the small-cluster-size region, $n_s(p)$ decays exponentially and then exhibits power-law behavior with $2 < \tau \leq 3$. Solid (black), dashed (red), and dashed-dotted (blue) lines represent $n_s(p)$ for $p=0.472576$, 0.657, and 0.65945, respectively. Two dotted lines are guide lines with slopes of $-2$ and $-3$. {\bf c}. For $p \geq p_c$, $n_s(p)$ for finite clusters shows exponentially decaying distributions. Solid curves represent $p=0.6596$, $0.7$, $0.8$, $0.9$, and $1.0$ from right to left. Dotted curve is an exponentially decaying guide curve.}
\label{fig:n_s_vs_s_at_any_p}
\end{figure*}

Using the rate equation approach and performing numerical simulations, we find the following phase transition properties: There exist two transition points $p_b$ and $p_c$ ($p_b < p_c$). The giant cluster size per node is zero asymptotically for $p < p_c$; jumps at $p_c$; and is finite for $p > p_c$. The size distribution of finite clusters $n_s$ decays in a power-law manner without any cutoff as $n_s\sim s^{-\tau(p)}$ for $p < p_c$. But it decays exponentially for $p > p_c$. Interestingly, the exponent $\tau(p) > 3$ for $p < p_b$ and $2 < \tau < 3$ in $p_b < p < p_c$. Thus the susceptibility, i.e., the mean cluster size $\sum s^2 n_s$, is finite and diverges in the former and the latter regions, respectively. Therefore, properties of an infinite-order transition, a second-order transition, and a first-order transition appear in the regions $p < p_b$, $p_b < p < p_c$, and $p_c < p$, respectively. The behaviors of the order parameter and the mean cluster size are depicted schematically in Figs.~1{\bf c} and 1{\bf d}, respectively. Moreover, the cluster size distributions in the three regimes are shown in Fig~\ref{fig:n_s_vs_s_at_any_p}. 

The pure power-law behavior of $n_s(p)\sim s^{-\tau(p)}$ for all $p < p_c$ is an intrinsic feature of the BKT transition. The exponent $\tau$ depends on $p$. At each time, a new node is added. Such a new node is not necessarily to merge into an existing cluster immediately but can be accumulated. Their population is comparable to others, so the frequency of merging between a single cluster and a finite cluster is also comparable to that between two finite clusters. When dynamics reaches a steady state, the cluster merging dynamics is self-organized and forms a power-law behavior of $n_s(p)$. As $p$ is increased, more links are added and the largest cluster becomes larger, and thus the exponent $\tau(p)$ is continuously decreasing. 

The underlying mechanism of the first-order transition is rather abnormal: Due to the suppression effect, the transition point is delayed, thus $\tau(p)$ can decrease down to two at $p_c$ in the $r$-GRN model. On the other hand, if the cluster size distribution follows a power law without any exponential cutoff, the largest cluster size scales with the current total number of nodes $N(t)$ in the steady state as $s_{\rm max} \sim N^{1/(\tau-1)}$. When $\tau$ decreases down to two, the largest cluster grows to the extent of the system size in the steady state. Even though the largest cluster size  increases continuously as $p$ is increased to $p_c$ in finite systems, the giant cluster size is subextensive to $N(t)$ for $p < p_c$ and it becomes extensive to $N(t)$ at $p=p_c$. Thus, a discontinuous transition occurs at $p_c$ in the thermodynamic limit.  

\begin{figure}[t]
\includegraphics[width=1.0\linewidth]{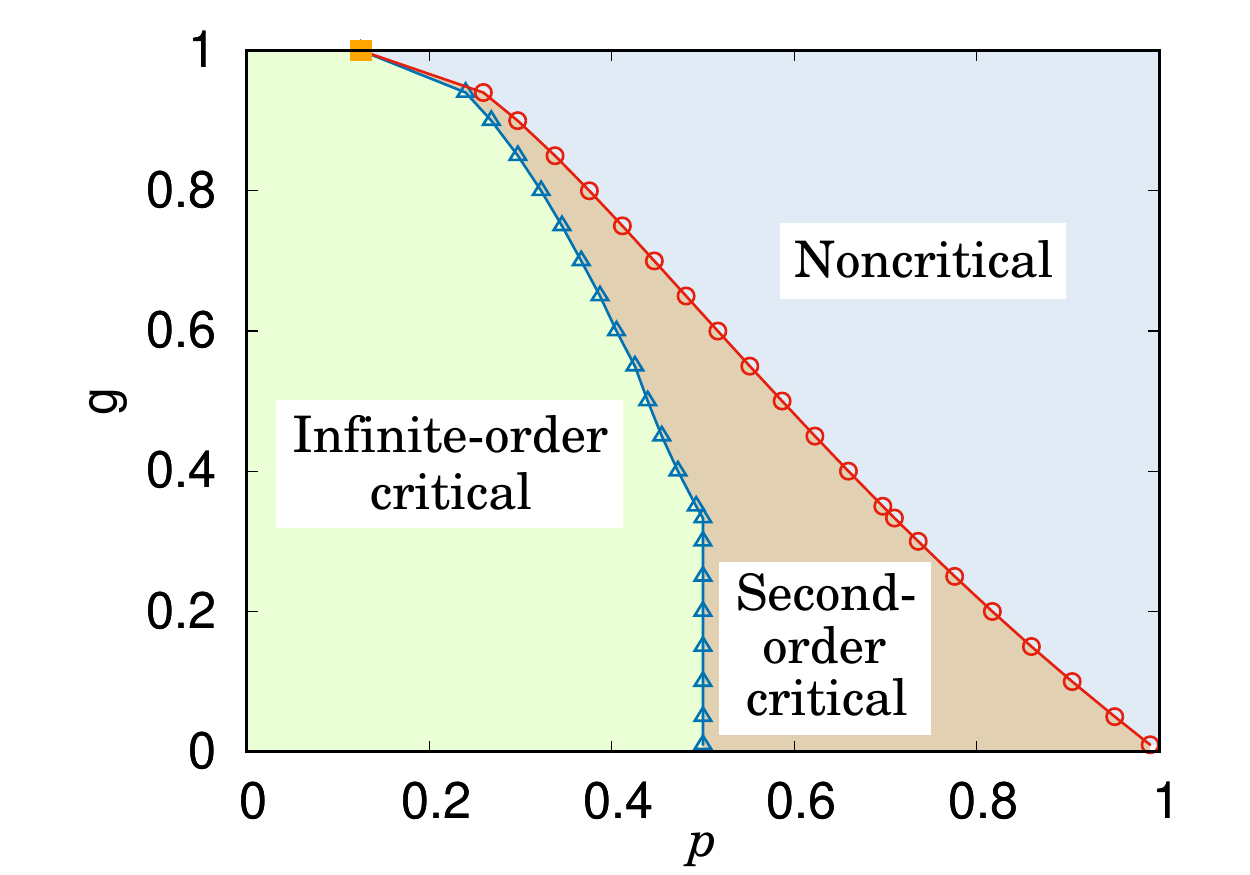}
\caption{{\bf Phase diagram of the $r$-GRN model.} Two transition points $p_{b}$ \textcolor{blue}{$\triangle$} and $p_{c}$ \textcolor{red}{$\bigcirc$} are determined for various $g$. $n_s(p,g)$ decays following a power law with $\tau > 3$ in the infinite-order critical region and $2 < \tau < 3$ in the second-order critical region. Thus, the mean cluster size is finite and diverges in those regions, respectively. As $g$ approaches one, the two transition points are closer and converge to the critical point of an infinite-order transition, represented by \textcolor{orange}{$\blacksquare$}.}
\label{fig:pc_vs_g}
\end{figure}

The exponent $\tau$ depends on the model parameter $g$, which is related to the suppression strength. $gN$  nodes in the small-cluster group have twice the chance to be linked, while the rest $(1-g)N$ nodes have one chance. When $g=1$, the $r$-GRN model reduces to the GRN model without any suppression effect; however, in the limit $g\to 0$, only isolated nodes have twice the chance, while the other nodes have only one chance. Thus the suppression strength becomes large as $g$ is decreased. Figure~\ref{fig:pc_vs_g} shows the phase diagram of the three phases as a function of the parameters $g$ and $p$. Indeed the phase boundaries determined by the criteria $\tau=3$ and $\tau=2$ depend on $g$. In the limit $g\to 0$, we derive $\tau=1+1/p$ explicitly, which is presented in the supplementary information (SI).    

We perform the global suppression dynamics for a different growing network model, the protein interaction model networks proposed by Sol\'e {\it et al.}~\cite{sole,vespignani,kim}. The Sol\'e model has a different evolution rule from that of the GRN model, but it exhibits a BKT-type percolation transition. We apply the global suppression dynamic rule to the Sol\'e model, and obtain a similar pattern of successive phase transitions as we obtained in the $r$-GRN model. The detailed results are presented in the SI. Thus, our main results are universal independent of detailed dynamic rules.

The BKT transition was found originally in the two-dimensional XY model in thermal systems~\cite{bere_1971,bere_1972,kt_1972,kt_1973,kt_1974}. The underlying mechanism of the thermal BKT transition is different from that of the growing percolation, but there exist some common properties: The singular part of the free energy of the XY model behaves as $f(t) \sim \exp({-bt^{-1/2}})$ with a positive constant $b$ for the reduced temperature $t=(T-T_c)/T_c >0$. The PT order parameter $G(p)$ of the GRN model behaves similarly for $p > p_c$. 

The correlation function decays in a power-law manner as $\Gamma(r)\sim r^{-\eta(T)}$ for $t < 0$ in the thermal system, where $\eta(T)\sim T$ continuously varies depending on temperature $T$. The correlation length $\xi=\infty$ for $t < 0$. This continuous varying exponent $\eta(T)$ corresponds to the exponent $\tau(p)-1$ in percolation, because the susceptibility is obtained from $\chi \sim \int d^2 r \Gamma(r)$ in the thermal system and $\sum s^2n_s$ in percolation. The characteristic size behaves as $s^*=\infty$ for $p < p_c$ in percolation, corresponding to $\xi(T)$ for $t < 0$ in the thermal system. While the susceptibility diverges for $\eta < 2$ in the thermal systems, it does for $\tau < 3$ in percolation system. For the GRN model proposed by Callaway et al.~\cite{Callaway}, $\tau > 3$ for $p \le p_c$ by a logarithmic correction at $p_c$~\cite{kim,mendes_BKT} and thus the susceptibility is finite at $p=p_c$. However, for the $r$-GRN model, it diverges in the region $p_b \le p \le p_c$, because $2\le \tau \le 3$ in that region. Thus, $\chi$ diverges at $p_c$. This result implies that the $r$-GRN model behaves more similarly to the BKT transition in the thermal system compared with the GRN model. 

The BKT transition can occur even in static networks. For instance, the percolation model in one-dimension with $1/r^2$ long-range connections~\cite{grassberger_BKT} and on hierarchical networks with short-range and long-range connections~\cite{berker_BKT,Boettcher} exhibit the BKT infinite-order transitions. It would be interesting to check whether or not the diverse phases and phase transitions we obtained occur in those static network models when the global suppression rule is applied. The pattern formation by topological defects in liquid crystals recently draws considerable attention~\cite{active_rmp,active_selinger}. Various patterns generated in that system are governed basically by the BKT transition. It would be non-trivial and interesting to note how those patterns can be changed when the system is subject to a certain global suppression dynamics. 

Having the exponent $\tau=2$ at $p_c$ reduced from $\tau=3$ by the global suppression effect implies that the system exhibits the maximum diversity of cluster sizes. In complex systems, diversity is a crucial factor to sustain a system in diverse fields such as finance-ecological systems~\cite{finan_eco} and evolving bio-systems~\cite{ecological}. Thus, the idea of the global suppression may be helpful for  establishing affirmative action policies in financial systems or other evolving systems. Moreover, this idea could be applied to diverse non-equilibrium phenomenon-based models such as epidemic models~\cite{epidemic}, voter models~\cite{voter} and so on.  \\

\noindent
\textsf{Methods:}
Let us introduce the dynamic rule of the $r$-GRN model. At the beginning, a system contain a single node. At each time, a new node is added to the system. Thus, the total number of nodes at time $t$ becomes $N(t)=t+1$. As time goes on, clusters of connected nodes form. We classify clusters into two sets, a set $R$ and its complementary set $R^c$, according to their sizes. The set $R$ contains $gN$ nodes belonging to the smallest clusters, while the set $R^c$ contains the rest $(1-g)N$ nodes. $g \in [0,1]$ is a parameter that controls the size of $R$. Rigorously speaking, let $s_i$ denote the size of the $i$-th cluster in ascending size order. If $\sum_{i = 1}^{k} s_i=\lceil gN \rceil$, then the set $R$ contains those $k$ smallest clusters. If $\sum_{i = 1}^{k-1} s_i < \lceil gN \rceil < \sum_{i = 1}^{k} s_i$, then $\lceil gN \rceil-\sum_{i = 1}^{k-1} s_i$ nodes are selected randomly from the $k$-th cluster and those nodes are included in the set $R$. The complementary set $R^c$ contains the nodes in the remaining (largest) clusters and the nodes left in the $k$-th cluster. Next, one node is selected randomly from the set $R$ and another is selected from among all the nodes. Then, a link is added between the two selected nodes with link occupation probability $p$.

\vskip 0.5cm

\noindent
\textbf{Acknowledgments}\\
This research was supported by the National Research Foundation of Korea (NRF) through Grant Nos. NRF-2014R1A3A2069005 (BK) and NRF-2017R1D1A1B03032864 (SWS), and a TJ Park Science Fellowship from the POSCO TJ Park Foundation (SWS).  


\noindent
\textbf{Competing interests}\\
The authors declare no competing interests.

\noindent
\textbf{Materials \& Correspondence} should be addressed to 
CCSS, CTP and Department of Physics and Astronomy, Seoul National University, Seoul 08826, Korea, bkahng@snu.ac.kr

\end{document}